\documentclass[article, shortnames, nojss, noheadings, notitle]{jss}
\usepackage{amsmath,amsfonts,bm}

\usepackage{thumbpdf}

\usepackage[colorinlistoftodos]{todonotes}

\usepackage{tikz}
\usetikzlibrary{arrows,matrix,backgrounds,shapes,positioning,fit,chains}

\usepackage{rotating}

\usepackage[noend]{algorithmic}
\usepackage{algorithm}

\usepackage{caption}

\usepackage{comment}

\pgfrealjobname{UsingSpikeSlabGAM}

\newcommand{\bvec}{\left[\begin{array}{c}}
\newcommand{\evec}{\end{array}\right]}
\newcommand{\bmat}[1]{\left[\begin{array}{*{#1}{c}}}
\newcommand{\emat}{\end{array}\right]}

\newcommand{\priorsim}{\stackrel{\text{prior}}{\sim}}
\newcommand{\ftilde}{\raise.17ex\hbox{$\scriptstyle\mathtt{\sim}$}}

\newcommand{\EW}{\operatorname{E}}

\newcommand{\iid} {\operatorname{i.i.d.}}

\title{\pkg{spikeSlabGAM}: Bayesian Variable Selection, Model Choice and Regularization for Generalized
    Additive Mixed Models in \proglang{R}}
\author{Fabian Scheipl \\ LMU M{\"u}nchen}

\Plainauthor{Fabian Scheipl}
\Plaintitle{spikeSlabGAM: Bayesian Variable Selection, Model Choice and Regularization for Generalized
    Additive Mixed Models in R} 
\Shorttitle{spikeSlabGAM: Variable Selection and Model Choice for GAMMs} 

\Abstract{
  The  \proglang{R} package \pkg{spikeSlabGAM} implements Bayesian variable selection, model choice, and regularized estimation in 
(geo-)additive mixed models for Gaussian, binomial, and Poisson responses. Its purpose is to (1) choose an
appropriate subset of potential covariates and their interactions, (2) to determine whether linear or more
flexible functional forms are required to model the effects of the respective covariates, and (3) to estimate
their shapes. Selection and regularization of the model terms is based on a novel spike-and-slab-type prior
on coefficient groups associated with parametric and semi-parametric effects. 
}
\Plainkeywords{MCMC, P-splines, spike-and-slab prior, normal-inverse-gamma}

\Address{
  Fabian Scheipl\\
  Institut f\"ur Statistik\\
  LMU M\"unchen\\
  Ludwigstr. 33\\
  80539 M\"unchen, Germany\\
  E-mail: \email{Fabian.Scheipl@stat.uni-muenchen.de}\\
  URL: \url{http:/www.stat.uni-muenchen.de/~scheipl/}
}
\begin{document}

\maketitle

\section{Introduction}

In data sets with many potential predictors, choosing an appropriate
subset of covariates and their interactions at the same time as
determining whether linear or more flexible functional forms are
required to model the relationships between covariates and the response
is a challenging and important task.   
From a Bayesian perspective, it
can be translated into a question of estimating marginal posterior
probabilities of whether a variable should be in the model and in what
form (i.e., linear or smooth; as a main effect and/or as an effect
modifier).

We introduce the \proglang{R} \citep{R} package \pkg{spikeSlabGAM} which implements fully Bayesian variable selection and
model choice with a \emph{spike-and-slab} prior 
structure that expands the approach in \citet{Ishwaran:2005} to select or deselect single
coefficients as well as blocks of coefficients associated with specific model terms.
The spike-and-slab priors we use are bimodal priors for the hyper-variances of the
regression coefficients which result in a two component mixture of a narrow spike around zero and a slab with wide support 
for the marginal prior of the coefficients themselves. 
The posterior mixture weights for the
spike component  for a specific coefficient or coefficient batch can be interpreted
as the posterior probability of its exclusion from the model.

The coefficient batches selected or deselected in this fashion can 
be associated with a wide variety of model terms such as simple linear terms,
factor variables, basis expansions for the modeling of smooth
curves or surfaces, intrinsically Gaussian Markov random fields (IGMRF), random effects, and
all their interactions.  \pkg{spikeSlabGAM} is able to deal with Gaussian, binomial and Poisson responses,
and can be used to fit piecewise exponential models for time-to-event data.  
For these response types, the package presented here implements regularized estimation, 
term selection, model choice, and model averaging for a similarly broad class of models as that 
available in \pkg{mboost} \citep{mboost} or \proglang{BayesX} \citep{bayesx}.
To the best of our knowledge, it is the first implementation of a 
Bayesian model term selection method that: 
(1) is able to fit models for non-Gaussian responses from the exponential family;
(2) selects and estimates many types of regularized effects with a
(conditionally) Gaussian prior such as simple covariates (both
metric and categorical), penalized splines (uni- or multivariate),
random effects, spatial effects (kriging, IGMRF) and their interactions; (3) and can 
distinguish between smooth nonlinear and linear effects.
The approach scales reasonably well to datasets with thousands
of observations and a few hundred coefficients and
is available in documented open source software.
%

Bayesian function selection, similar to the frequentist COSSO procedure
\citep{Lin:Zhang:2006}, is usually based on decomposing the additive
model in the spirit of a smoothing spline
ANOVA \citep{Wahba:Wang:Gu:1995}. \citet{Wood:Kohn:2002} and
\citet{Yau:Kohn:Wood:2003} describe procedures for Gaussian and 
latent Gaussian models using a
data-based prior that requires two MCMC runs, a pilot run to obtain
a data-based prior for the slab part and a second one to
estimate parameters and select model components. A more general
approach that also allows for flexible modeling of the dispersion in double exponential regression
models is described in \citet{Cottet:Kohn:Nott:2008}, but 
no implementation is available.
\citet{Reich:Storlie:Bondell:2009} also use the smoothing spline
ANOVA framework and perform variable and function selection via SSVS
for Gaussian responses. \citet{Fruehwirt:Wagner:2010} discuss various spike-and-slab 
prior variants for the selection of random intercepts for Gaussian and 
latent Gaussian models. 


The remainder of this paper is structured as follows: 
Section \ref{background} gives some background on the two main ideas used in \pkg{spikeSlabGAM}.
\ref{gamm} introduces the necessary notation for 
the generalized additive mixed model and \ref{spikeSlabSSVS} fills in some details on the spike-and-slab prior.
Section \ref{implementation} relates
details of the implementation: how the design matrices for the model terms are constructed (Section \ref{design}) and 
how the MCMC sampler works (Section \ref{mcmc}). Section \ref{using} explains how to specify, visualize
and interpret models fitted with \pkg{spikeSlabGAM} and contains an application to the Pima Indian Diabetes dataset.

\newpage

\section{Background}\label{background}
\subsection{Generalized additive mixed models}\label{gamm}

The generalized additive mixed model (GAMM) is a broad model class
that forms a subset of structured additive regression
\citep{Fahr:Kneib:Lang:2004}. In a GAMM, the distribution of the responses $\bm{y}$ given a set of covariates $\bm x_j\;
(j=1,\dots,p)$ belongs to an exponential family, i.e.,
\begin{align*}
    \pi(y |x, \phi) &= c(y, \phi)\exp\left(\frac{
    y \theta- b(\theta)}{\phi}\right),
\end{align*}
with $\theta$,$\phi$,$b(\cdot)$ and $c(\cdot)$ determined by the type of
distribution. 
The conditional expected value of the response $\EW(\bm y | \bm x_1,\ldots, \bm x_p) = h(\bm \eta)$ 
is determined by the additive predictor $\bm \eta$ and a fixed response function $h(\cdot)$.

The additive predictor
\begin{align}\label{F:eta} 
\bm\eta &= \bm\eta_o + \bm X_u \bm\beta_u + \sum^p_{j=1}f_j(\bm x) 
\end{align}
has three parts: a fixed and known offset $\bm\eta_o$, 
a linear predictor $\bm X_u \bm\beta_u$ for model terms that are not under selection with coefficients $\bm\beta_u$ 
associated with a very flat Gaussian prior (this will typically include at least a global intercept term), and the model terms 
$f_j(\bm x)=\left(f_j(\bm x_{1}),\ldots,f_j(\bm x_{n})\right)^\top\;(j=1,\dots,p)$ that are each represented as linear combinations of $d_j$ basis functions $B_j(\cdot)$ so that
\begin{align}
\begin{split}
f_j(\bm x) &= \sum_{k=1}^{d_j} \beta_{jk}B_{jk}(\bm x)= \bm B_j \bm \beta_j,  \text{ with } B_{jk}(\bm x)=\left(B_{jk}(\bm x_{1}),\ldots,B_{jk}(\bm x_{n})^\top\right)\\
\text{ and }\bm\beta_j &\priorsim \operatorname{peNMIG}(v_0, w, a_\tau, b_\tau) \text{ for } j=1,\dots,p. \label{F:term}
\end{split} 
\end{align}
The peNMIG prior structure is explained in detail in Section \ref{spikeSlabSSVS}.

Components $f_j(\bm x)$ of the additive predictor represent a wide variety of model terms, such as
(1) linear terms ($f_j(\bm x) = \beta_j \bm x_j$),
(2) nominal or ordinal covariates ($f(x_{ji}) = \beta_{x(k)}$ iff $x_{ji} = k$, i.e., if entry $i$ in $\bm x_j$ is $k$),
(3) smooth functions of (one or more) continuous covariates (splines, kriging effects, tensor product splines or varying coefficient terms, e.g., \citet{Wood:2006}),
(4) Markov random fields for discrete spatial covariates \citep[e.g.][]{Rue:Held:2005}, 
(5) random effects (subject-specific intercepts or slope), and (6) interactions between the different terms (varying-coefficient models, effect modifiers, factor interactions).
Estimates for semiparametric model terms and random effects are regularized in order to
avoid overfitting and modeled with appropriate shrinkage priors.
These shrinkage or regularization priors are usually Gaussian or can
be parameterized as scale mixtures of Gaussians \citep[e.g.][]{Fahr:Kneib:Konrath:2010}.
The peNMIG variable selection prior used in \pkg{spikeSlabGAM} can also be viewed as a scale mixture of Gaussians.
  
\newpage 
\subsection{Stochastic search variable selection and spike-and-slab priors}\label{spikeSlabSSVS}
While analyses can benefit immensely from a flexible and versatile array of potential model terms, 
the large number of possible models in any given data situation calls for a principled procedure
that is able to select the covariates that
are relevant for the modeling effort (i.e., variable selection) as well as to determine
the shapes of their effects (e.g., smooth vs. linear) and which interaction effects or effect modifiers
need to be considered (i.e., model choice). SSVS and spike-and-slab priors are Bayesian methods
for these tasks that do not rely on the often very difficult calculation of marginal likelihoods for 
large collections of complex models \citep[e.g.][]{Han:Carlin:2001}. 
     
The basic idea of the SSVS approach \citep{George:McCulloch:1993} is to introduce a binary latent variable 
$\gamma_j$ associated with the coefficients $\bm{\beta}_j$ of each model term so that the contribution of a model
term to the predictor is forced to be zero -- or at least negligibly small -- if $\gamma_j$ is in one state and 
left unchanged if $\gamma_j$ is in the other state. 
The posterior distribution of $\gamma_j$ can be interpreted as marginal 
posterior probabilities for exclusion or inclusion of the respective model term. The posterior distribution 
of the vector $\bm\gamma = (\gamma_1,\ldots,\gamma_p)^\top$ can be interpreted as posterior probabilities for the different
models represented by the configurations of $\bm\gamma$. Another way to express this basic idea is to assume 
a spike-and-slab mixture prior for each $\bm{\beta}_j$, with one component being a narrow spike around the origin that
imposes very strong shrinkage on the coefficients and the other component being a wide slab that imposes
very little shrinkage on the coefficients. The posterior weights for the spike and the slab can then be interpreted 
analogously. 

The flavor of spike-and-slab prior used in \pkg{spikeSlabGAM} is a further development based on \citet{Ishwaran:2005}:
The basic prior structure, 
which we call a Normal - mixture of inverse Gammas (NMIG) prior, uses
a bimodal prior on the variance $v^2$ of the coefficients that results in a spike-and-slab type prior on 
the coefficients themselves. For a scalar $\beta$, the prior structure is given by: 
\begin{align}\label{F:basicNMIG}
\begin{split}
\beta|\gamma, \tau^2 &\priorsim N(0, v^2)\text{ with }v^2 = \tau^2\gamma,\\
\gamma|w &\priorsim w I_1(\gamma) + (1-w)I_{v_0}(\gamma), \\
\tau^2 &\priorsim \Gamma^{-1}(a_\tau, b_\tau), \\
\text{and } w &\priorsim \operatorname{Beta}(a_w, b_w).
\end{split}
\end{align}
$I_x(y)$ denotes a function that is 1 in $x$ and 0 everywhere else and $v_0$ is some small positive constant,
so that the indicator $\gamma$ is 1 with probability $w$ and close to zero with probability $1-w$.
This means that the effective prior variance $v^2$ is very small if $\gamma=v_0$ 
--- this is the spike part of the prior. The variance $\tau^2$ is sampled from an informative 
Inverse Gamma ($\Gamma^{-1}$) prior with density $p(x| a, b) = \tfrac{b^a}{\Gamma(a)} x^{(a-1)} \exp\left(-\tfrac{b}{x}\right)$.

This prior hierarchy has some advantages for selection of model terms for non-Gaussian data
since the selection (i.e., the sampling of indicator variables
$\gamma$) occurs on the level of the coefficient variance. This means that the likelihood itself is not in the
Markov blanket of $\gamma$ and consequently does not occur in the
full conditional densities (FCD) for the indicator variables, so that the FCD for $\gamma$ is available in closed form
regardless of the likelihood. However,
since only the regression coefficients and not the data itself occur in the Markov blanket
of $\bm\gamma$, inclusion or exclusion of model terms is based on the magnitude of the coefficients $\bm \beta$ and
not on the magnitude 
of the effect $\bm{B \beta}$ itself. This means that design matrices have to be scaled similarly
across all model terms for the magnitude of the coefficients to be a good proxy for the importance of the associated effect.
In \pkg{spikeSlabGAM}, each term's design matrix is scaled to have a Frobenius norm of 0.5 to achieve this.

\subsubsection*{A parameter-expanded NMIG prior}     
While the conventional NMIG prior \eqref{F:basicNMIG} works well for the selection of single coefficients, 
it is unsuited for the simultaneous selection or deselection of coefficient vectors, such as coefficients associated with 
spline basis functions or with the levels of a random intercept. In a nutshell,
the problem is that a small variance for a batch of
coefficients implies small coefficient values and small coefficient
values in turn imply a small variance so that blockwise MCMC samplers are
unlikely to exit a basin of attraction around the origin. \citet{Gelman:2008}
analyze this issue in the context of hierarchical models, where
it is framed as a problematically strong dependence between a block of
coefficients and their associated hypervariance. A bimodal prior for
the variance, such as the NMIG prior, obviously exacerbates these
difficulties as the chain has to be able to switch between the different components of the mixture prior.
The problem is much less acute for coefficient batches with only a single or few entries since
a small batch contributes much less information to the full conditional of its variance parameter. 
The sampler is then better able to switch between the less clearly separated basins of attraction around 
the two modes corresponding to the spike and the slab \citep[Section 3.2]{Scheipl:2010}. 
In our context, ``switching modes'' means that 
entries in $\bm\gamma$ change their state from 1 to $v_0$ or vice versa.
The practical importance for our aim is clear: Without fast and reliable mixing
of $\bm\gamma$ for coefficient batches with more than a few entries, the posterior distribution cannot be used to define marginal probabilities of models or
term inclusion.
In previous approaches, this problem has been circumvented by either relying on very low dimensional bases with only 
a handful of basis functions \citep{Reich:Storlie:Bondell:2009,Cottet:Kohn:Nott:2008} or by sampling the indicators from a 
partial conditional density, with coefficients and their hypervariances integrated out \citep{Yau:Kohn:Wood:2003}. 

A promising strategy to reduce the dependence between coefficient batches 
and their variance parameter that neither limits the dimension of the base nor
relies on repeated integration of multivariate functions is the
introduction of working parameters that are only partially
identifiable along the lines of \emph{parameter expansion} or
\emph{marginal augmentation} \citep{Meng:Dyk:1997, Gelman:2008}. 
The central idea implemented in \pkg{spikeSlabGAM} is  
a multiplicative parameter expansion 
that improves the shrinkage properties of the resulting
marginal prior compared to NMIG \citep[Section 4.1.1]{Scheipl:2011} and
enables simultaneous selection or deselection of large coefficient
batches. 

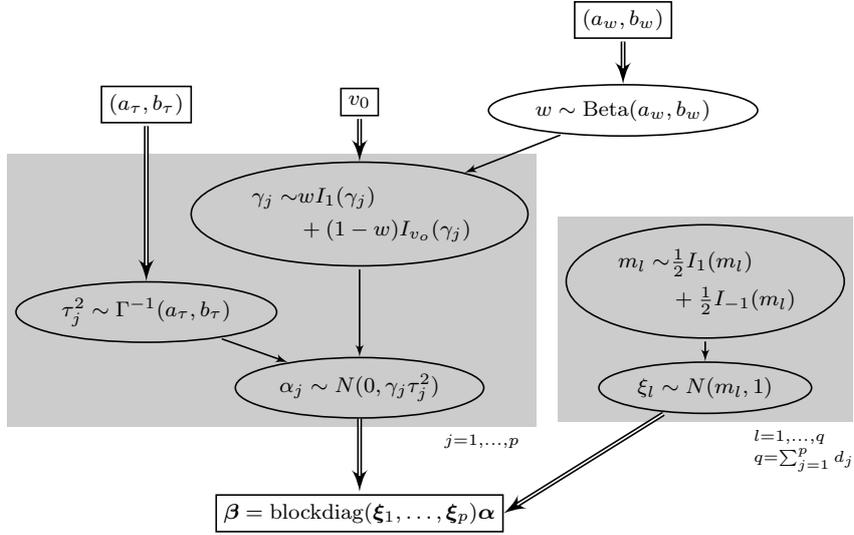
\begin{figure}[!htp]
\begin{center}
\begin{scriptsize}
\textbf{peNMIG: Normal mixture of inverse Gammas with parameter expansion}\\
$\phantom{bla}$\\
  \begin{tikzpicture}
[hyper/.style={rectangle, draw, semithick},
 param/.style={ellipse, draw, semithick},
 parent/.style = {<-, shorten <=1pt, >=latex', semithick},
 child/.style = {->, shorten <=1pt, >=latex', semithick},
 determparent/.style = {<-, double, shorten <=1pt, >=latex', semithick},
 determchild/.style = {->, double, shorten <=1pt, >=latex', semithick}] 
    \node[param](alpha)  {$\alpha_j \sim N(0, \gamma_j\tau_j^2)$};
    \node[param](gamma)  [above =4em of alpha] {$\begin{aligned}\gamma_j \sim &  w  I_1(\gamma_j) \\& {} + (1-w)I_{v_o}(\gamma_j)\end{aligned}$}
        edge[child] (alpha);
    \node[hyper](v0) [above =2em of gamma]    {$v_0$}
        edge[determchild] (gamma);
    \node[param](w)  [above right=3em of gamma] {$w \sim \operatorname{Beta}(a_w, b_w)$}
        edge[child] (gamma);
    \node[hyper](hyperw)  [above =2em of w]    {$(a_w, b_w)$}
        edge[determchild] (w);
    \node[param](tau)  [above left=2em of alpha] {$\tau_j^2 \sim \Gamma^{-1}(a_\tau, b_\tau)$}
        edge[child] (alpha);
    \node[hyper](hypertau)   [above=7em of tau]  {$(a_\tau, b_\tau)$}
        edge[determchild] (tau);
     \begin{pgfonlayer}{background}
    \node[rectangle, fill=gray!40, fit = (alpha) (tau) (gamma) ,  label=-40:$^{j = 1,\dots,p}$]{};
    \end{pgfonlayer}
    \node[hyper](beta)  [below= of alpha]    {\mbox{$\bm \beta = \operatorname{blockdiag}(\bm
    \xi_1,\dots,\bm\xi_p) \bm \alpha$}}
        edge[determparent]    (alpha);
    \node[param](xi)    [right=5em of alpha]    {$\xi_l \sim N(m_l, 1)$}
     edge[determchild]    (beta.east);
    \node[param](m)     [above=1em of xi]       {$\begin{aligned} m_l
    \sim &
    \tfrac{1}{2}I_{1}(m_l)\\ & {} + \tfrac{1}{2}I_{-1}(m_l)\end{aligned}$} edge[child] (xi);
    \begin{pgfonlayer}{background}
    \node[rectangle, fill=gray!40, fit = (xi) (m), label=-90:$\phantom{lkjadghkljghlkjdhg}^{l =
    1,\dots,q}_{q=\sum_{j=1}^p d_j}$](fitxim){};
    \end{pgfonlayer}
\end{tikzpicture}
  \caption[DAG of peNMIG prior]{Directed acyclic graph of the peNMIG
  prior structure. Ellipses are stochastic nodes, rectangles are deterministic nodes.
  Single arrows are stochastic edges, double arrows are deterministic
  edges.}
  \label{parExpDAG}
\end{scriptsize} 
\end{center}
\end{figure}
Figure~\ref{parExpDAG} shows the peNMIG
prior hierarchy for a model with $p$ model terms: We set
$\bm \beta_j = \alpha_j \bm\xi_j$ with mutually independent $\alpha_j$ and $\bm
\xi_j$ for a coefficient batch $\bm\beta_j$ with
length $d_j$ and use a scalar parameter $\alpha_j \priorsim
\operatorname{NMIG}(v_0, w, a_\tau, b_\tau)$, where NMIG
denotes the prior hierarchy given in \eqref{F:basicNMIG}. Entries of
the vector $\bm\xi_j$ are $\iid$ \mbox{$\xi_{jk}
\priorsim N(m_{jk},1)\;(k=1,\dots,d_j; j=1,\dots,p)$} with prior means $m_{jk}$ either 1 or -1 with equal probability. 

We write 
\begin{align}\label{F:peNMIG}
\bm\beta_j &\sim \operatorname{peNMIG}(v_0, w, a_\tau, b_\tau)
\end{align}
as shorthand for this parameter expanded NMIG prior. 
The effective dimension of each coefficient batch associated with
a specific $\gamma_j$ and $\tau^2_j$ is then just one, 
since the Markov blankets of both $\gamma_j$ and
$\tau_j$ now only contain the scalar parameter $\alpha_j$ instead of
the vector $\bm \beta_j$. This solves the
mixing problems for $\bm\gamma$ described above.  
The long vector $\bm \xi=(\bm \xi_1^\top, \dots,\bm \xi_p^\top)^\top$ is
decomposed into subvectors $\bm \xi_j$ associated with the different
coefficient batches and their respective entries $\alpha_j\;(j=1,\ldots,p)$ in $\bm\alpha$. The parameter $w$ is a global parameter that 
influences all model terms, it can be interpreted as the prior probability of a 
term being included in the model. The parameter $\alpha_j$ parameterizes
the ``importance'' of the $j$-th model term, while $\bm\xi_j$
``distributes'' $\alpha_j$ across the entries in the coefficient batch $\bm \beta_j$.
Setting the conditional expectation $\EW(\xi_{jk}|m_{jk}) = \pm 1$ shrinks $|\xi_{jk}|$ towards $1$, the
multiplicative identity, so that the interpretation of $\alpha_j$ as
the ``importance'' of the $j$-th coefficient batch can be maintained.

The marginal peNMIG prior, i.e., the prior for $\bm\beta$ integrated over
the intermediate quantities $\alpha$, $\bm \xi$, $\tau^2, \gamma$ and $w$,
combines an infinite spike at zero with heavy
tails. This desirable combination is similar to the properties of 
other recently proposed shrinkage
priors such as the horseshoe prior \citep{Carvalho:2010} and the
normal-Jeffreys prior \citep{Bae:Mallick:2004} for which both
robustness for large values of $\beta$ and very efficient estimation
of sparse coefficient vectors have been shown 
\citep{Polson:Scott:2010}. 
The shape of the marginal peNMIG prior is fairly close to the 
original spike-and-slab prior suggested by
\citet{Mitchell:Beauchamp:1988}, which used a mixture of a point
mass in zero and a uniform distribution on a finite interval, but
it has the benefit of (partially) conjugate and proper
priors. A detailed derivation of the properties of the peNMIG prior 
and an investigation of its sensitivity to hyperparameter settings
is in \citet{Scheipl:2011}, along with
performance comparisons against \pkg{mboost}
and other approaches with regard to 
term selection, sparsity recovery, and 
estimation error for Gaussian, binomial and Poisson responses on
real and simulated data sets.  
The default settings for the hyperparameters, validated through many simulations and
data examples are $a_\tau=5, b_\tau=25, v_0=2.5\cdot10^{-4}$. By default, 
we use a uniform prior on $w$, i.e., $a_w=b_w=1$, and a very flat $\Gamma^{-1}(10^{-4},10^{-4})$ prior
for the error variance in Gaussian models.

\section{Implementation}\label{implementation}

\subsection{Setting up the design}\label{design}
All of the terms implemented in \pkg{spikeSlabGAM} have the following structure: 
First, their contribution to the predictor $\bm \eta$ is represented as a linear combination of basis functions, 
i.e., the
term associated with covariate $\bm x$ is represented as 
$\tilde f(\bm x) = \sum_{k=1}^K \delta_{k}\tilde B_k(\bm x)= \bm{\tilde{B}} \bm \delta$,
where $\bm\delta$ is a vector of coefficients associated with
the basis functions $\tilde B_k(\cdot)\;( k=1,\ldots,K)$ evaluated in $\bm x$.
Second, $\bm\delta$ has a (conditionally) multivariate Gaussian prior, i.e., $\bm\delta|v^2 \priorsim N(\bm 0, v^2\bm P^-)$,
with a fixed scaled precision matrix $\bm P$ that is often positive \emph{semi}-definite. 
Table \ref{T:termtypes} gives an overview of the model terms available in \pkg{spikeSlabGAM} and how they fit into this
framework.
\begin{table}[!tbp]
\begin{small}
\begin{center}
\begin{tabular}{l|p{.25\textwidth}|p{.25\textwidth}|p{.25\textwidth}}
\textbf{\proglang{R}-syntax} & \textbf{Description} & $\bm{\tilde{B}}$ & $\bm P$\\
\hline
\code{lin(x, degree)} & linear/polynomial trend: basis functions are orthogonal polynomials of degree 1 to \code{degree} evaluated in \code{x}; defaults to \code{degree}$=1$& \code{poly(x, degree)} & identity matrix\\
\hline
\code{fct(x)} & factor: defaults to sum-to-zero contrasts & depends on contrasts & identity matrix\\
\hline  
\code{rnd(x, C)} & random intercept: defaults to $\iid$; i.e., correlation $\mathtt{C} = \bm I$ & indicator variables for each level of \code{x} & $\mathtt{C}^{-1}$\\
\hline
\code{sm(x)} & univariate penalized spline: defaults to cubic B-splines with $2^{\text{nd}}$ order difference penalty & B-spline basis functions  & ${\bm\Delta^{d}}^\top\bm\Delta^{d}$ with 
    $\bm\Delta^{d}$ the $d^{th}$ diff. operator matrix\\
\hline  
\code{srf(xy)} & penalized surface estimation on 2-D coordinates \code{xy}: 
defaults to tensor product cubic B-spline with first order difference penalties& (radial) basis functions (thin plate / tensor product B-spline) & depends on basis function\\
\hline  
\code{mrf(x, N)} & first order intrinsic Gauss-Markov random field: factor \code{x} defines 
the grouping of observations, \code{N} defines the neighborhood structure of the levels in \code{x} & indicator variables for regions in \code{x} & precision matrix of MRF defined by (weighted) adjacency matrix \code{N}\\  
\end{tabular}
\end{center}
\caption[Term types]{Term types in \pkg{spikeSlabGAM}.  
The semiparametric terms (\code{sm()}, \code{srf()}, \code{mrf()}) only parameterize the proper part
of their respective regularization priors (see Section \ref{design}). Unpenalized terms not associated with a 
peNMIG prior (i.e., the columns in $\bm X_u$ in \eqref{F:eta}) are specified with term type \code{u()}.}
\label{T:termtypes}
\end{small}
\end{table}

Formula \eqref{F:term} glosses over the fact that every coefficient batch associated with a specific term will 
have some kind of prior dependency structure determined by $\bm P$. 
Moreover, if $\bm P$ is only positive \emph{semi}-definite, the prior is partially improper. For example, 
the precision matrix for a B-spline with second order difference penalty implies an improper flat prior on the 
linear and constant components of the estimated function \citep{Lang:Brezger:2004}. 
The precision matrix for an IGMRF of first order puts an improper flat prior on the mean level of the IGMRF \citep[][ch. 3]{Rue:Held:2005}.  
These partially improper priors for splines and IGMRFs are problematic for \pkg{spikeSlabGAM}'s purpose for two reasons:
In the first place, if e.g., coefficient vectors that parameterize linear functions are in the nullspace of the prior precision matrix, 
the linear component of the function is estimated entirely 
unpenalized. This means that it is unaffected by the variable selection property of the peNMIG prior and thus always remains 
included in the model, but we need to be able to not only remove the entire effect 
of a covariate (i.e., both its penalized and unpenalized parts) from the model, but also be able to select or deselect its
penalized and unpenalized parts separately.
The second issue is that, since the nullspaces of these precision matrices usually also contain coefficient vectors that 
parameterize constant effects, terms in multivariate models are not identifiable, since adding a constant to one term 
and subtracting it from another does not affect the posterior. 
  
Two strategies to resolve these issues are implemented in \pkg{spikeSlabGAM}. Both
involve two steps: (1) Splitting terms with partially improper priors into two parts -- one associated with 
the improper/unpenalized part of the prior and one associated with the proper/penalized part of the prior; 
and (2) absorbing the fixed prior correlation
structure of the coefficients implied by $\bm P$ into a transformed design matrix $\bm B$ associated with then
\emph{a priori} independent coefficients $\bm\beta$ for the penalized part. 
Constant functions contained in the unpenalized part of a term are subsumed into a global intercept.
This removes the identifiability issue. 
The remainder of the unpenalized component enters the model in a separate term, e.g.,   
P-splines (term type \code{sm()}, see Table \ref{T:termtypes}) leave polynomial functions of a certain order
unpenalized and these enter the model in a separate \code{lin()}-term. 

\paragraph*{Orthogonal decomposition}
The first strategy, used by default, 
employs a reduced rank approximation of the implied covariance of $\tilde f(\bm x)$ 
to construct $\bm B$, similar to the approaches used in \citet{Reich:Storlie:Bondell:2009} and \citet{Cottet:Kohn:Nott:2008}:\\
Since $\tilde f(\bm x) =  \bm{\tilde{B}}\delta \priorsim N(0, v^2 \bm{\tilde{B}P^-\tilde{B}^\top})$,
we can use the spectral decomposition $\bm{\tilde{B}P^-\tilde{B}^\top}= \bm{UDU}^\top$
with orthonormal $\bm U$ and diagonal $\bm D$  with entries $\geq 0$ to find an orthogonal basis
representation for $\COV\left( \tilde f(\bm x)\right)$. For $\bm{\tilde{B}}$ 
with $\tilde d$ columns and full column rank and $\bm P$ with rank $\tilde d - n_P$, where $n_P$ is the dimension 
of the nullspace of $\bm P$, all eigenvalues of $\COV\left(\tilde f(\bm x)\right)$ except the first $\tilde d - n_P$ 
are zero. Now write $\COV\left(\tilde f(\bm x)\right) = [\bm U_+  \bm U_0] \left[\begin{smallmatrix}  \bm D_+ & \bm 0 \\ \bm
 0 & \bm 0\end{smallmatrix}\right] [\bm U_+  \bm U_0]^\top$, where $\bm U_+$ is a matrix of eigenvectors associated
with the positive eigenvalues in $\bm D_+$, and $\bm U_0$ are the eigenvectors associated with the zero eigenvalues.
With $\bm B = \bm U_+ \bm D_+^{1/2}$ and $\bm\beta \priorsim N(0, v^2 \bm I)$, $f(\bm x)= \bm{ B  \beta}$ 
has a proper Gaussian distribution that is proportional to that of the partially improper prior of $\tilde f(\bm x)$ \citep[][eq. (3.16)]{Rue:Held:2005}
and parameterizes only the penalized/proper part of $\tilde f(\bm x)$, while the unpenalized part of the function is represented
by $\bm U_0$.

In practice, it is unnecessary and impractically slow to compute all $n$ eigenvectors and values 
for a full spectral decomposition $\bm{UDU}^\top$. 
Only the first $\tilde d - n_P$ are needed for $\bm B$, and of those the first few typically represent
most of the variability in $f(\bm x)$.  
\pkg{spikeSlabGAM} makes use of a fast truncated bidiagonalization algorithm \citep{Baglama:Reichel:2006}
implemented in \pkg{irlba} \citep{irlba} to compute only the largest $\tilde d - n_P$ eigenvalues
of $\COV\left(\tilde f(\bm x)\right)$ and their associated eigenvectors. Only the first 
$d$ eigenvectors and -values whose
sum represents at least $.995$ of the sum of all eigenvalues
are used to construct the reduced rank orthogonal basis $\bm B$ with $d$ columns. 
e.g., for a cubic P-spline with second order difference penalty and 20 basis functions       
(i.e., $\tilde d = 20$ columns in $\bm{\tilde{B}}$ and $n_P=2$), 
$\bm B$ will typically have only 8 to 12 columns.

\paragraph*{``Mixed model'' decomposition}
The second strategy reparameterizes via a decomposition of the coefficient
vector $\bm\delta$ into an \textbf{u}npenalized part and a \textbf{p}enalized part: 
$\bm \delta= \bm X_u \bm\beta_u + \bm X_p \bm
\beta$, where $\bm X_u$ is a
basis of the $n_P$-dimensional nullspace of $\bm P$ and $\bm X_p$ is 
a basis of its complement. \pkg{spikeSlabGAM} uses a spectral
decomposition of $\bm P$ with
 $\bm P = [\bm \Lambda_+ \bm \Lambda_0] \left[\begin{smallmatrix} \bm \Gamma_+ & \bm 0 \\ \bm
 0 & \bm 0\end{smallmatrix}\right] [\bm \Lambda_+  \bm \Lambda_0]^\top$,
where $\bm \Lambda_+$ is the matrix of eigenvectors associated with the
positive eigenvalues in $\bm \Gamma_+$, and $\bm \Lambda_0$ are the
eigenvectors associated with the zero eigenvalues.  This decomposition yields
$\bm X_u = \bm \Lambda_0$  and $\bm X_p = \bm L(\bm L^\top \bm L)^{-1}$
with $\bm{L} = \bm \Lambda_+ \bm{\Gamma}_+^{1/2}$. The model term can then be expressed as
$ \bm{\tilde{B}\delta} = \bm{\tilde{B}(\bm X_u \bm\beta_u + \bm X_p \bm \beta)=\bm{B}_u \beta_u} + \bm{B
\beta}$ 
with $\bm{B}_u$ as the design matrix associated with the unpenalized
part and $\bm B$ as the design matrix associated with the
pena\-lized part of the term. The prior for the coefficients associated with the penalized part
after re\-pa\-ra\-me\-teriza\-tion is then $\bm \beta \sim
N(\bm 0, v^2 \bm I)$, while $\bm \beta_u$ has a flat prior \citep[c.f.][ch. 5.1]{Kneib:06}. 

\paragraph*{Interactions}
Design matrices for interaction effects are constructed from tensor products 
(i.e., column-wise Kronecker products)
of the bases for the respective main effect terms.  
For example, the complete interaction between two numeric
covariates $x_1$ and $x_2$ with smooth effects modeled as P-splines with second order difference penalty 
consists of the interactions of their unpenalized parts (i.e., linear $x_1$-linear $x_2$), 
two varying-coefficient terms (i.e., smooth $x_1 \times$ linear $x_2$, linear $x_1 \times$ smooth $x_2$) and a 
2-D nonlinear effect (i.e., smooth $x_1 \times$ smooth $x_2$). 
By default, \pkg{spikeSlabGAM} uses a reduced rank representation of these tensor product bases derived from  
their partial singular value decomposition as described above for the ```orthogonal'' decomposition. 

\paragraph*{``Centering'' the effects}
By default, \pkg{spikeSlabGAM} makes the estimated effects of all terms orthogonal to the nullspace of their associated penalty
and, for interaction terms, against the corresponding main effects as in \citet{Yau:Kohn:Wood:2003}.
Every $\bm B$ is transformed via  
$\bm B \rightarrow \bm B \left(\bm I - \bm Z(\bm Z^\top \bm Z)^{-1} \bm
Z^\top\right)$. For simple terms (i.e., \code{fct()}, \code{lin()}, \code{rnd()}),
$\bm Z = \bm 1$ and the projection above simply enforces a sum-to-zero constraint
on the estimated effect. For semi-parametric terms, $\bm Z$ is a basis of the nullspace of the implied 
prior on the effect. For interactions between $d$ main effects, $\bm Z = \left[\bm 1\; \bm B_1\; \bm B_2 \ldots \bm B_d\right]$,
where $\bm B_1,\dots, \bm B_d$ are the design matrices of the involved main effects.
This centering improves separability between main effects and their interactions by removing any overlap of their respective column 
spaces. All uncertainty about the mean response level is shifted into the global
intercept. The projection uses the QR decomposition of $\bm Z$ for speed and stability.

\subsection{Markov chain Monte Carlo implementation}\label{mcmc}

\begin{algorithm}[!h]
\begin{algorithmic}
\STATE Initialize $\bm\tau^{2(0)}, \bm\gamma^{(0)},
\phi^{(0)}, w^{(0)}$ and $\bm\beta^{(0)}$ (via IWLS for
non-Gaussian response) \STATE
Compute $\bm\alpha^{(0)}, \bm\xi^{(0)}, \bm X^{(0)}_{\alpha}$
\FOR{iterations $t=1,\dots,T$}
    \FOR{blocks $b=1,\dots,b_\alpha$}
        \STATE update $\bm\alpha^{(t)}_b$ from its FCD (Gaussian case, see \eqref{F:FCDCoef})/ via P-IWLS
    \ENDFOR
    \STATE set $\bm X^{(t)}_\xi = \bm X \operatorname{blockdiag}(\bm 1_{d_1},\dots,\bm 1_{d_p})\bm\alpha^{(t)}$
    \STATE update $m^{(t)}_1,...,m^{(t)}_q$ from their FCD: $P(m^{(t)}_l=1|\cdot)=\tfrac{1}{1+\exp(-2\xi^{(t)}_l)}$
    \FOR{blocks $b=1,\dots,b_\xi$}
        \STATE update $\bm\xi^{(t)}_b$ from its FCD (Gaussian case, see \eqref{F:FCDCoef})/ via P-IWLS
    \ENDFOR
    \FOR{model terms $j=1,\dots,p$}
        \STATE rescale $\bm\xi^{(t)}_j$ and $\alpha^{(t)}_j$
    \ENDFOR
    \STATE set $\bm X^{(t)}_\alpha = \bm X \operatorname{blockdiag}(\bm\xi^{(t)}_1,\dots,\bm\xi^{(t)}_p)$
    \STATE update ${\tau_1}^{2(t)},...,{\tau_p}^{2(t)}$ from their FCD:
    $\tau^{2(t)}_j|\cdot \sim \Gamma^{-1}\left(a_\tau+ 1/2,  b_\tau + \frac{\alpha_{j}^{2(t)}}{2\gamma^{(t)}_j}\right)$ 
    \STATE update ${\gamma_1}^{(t)},...,{\gamma_p}^{(t)}$ from their FCD:
    $\frac{P(\gamma^{(t)}_j=1|\cdot)}{P(\gamma^{(t)}_j=v_0|\cdot)} =  v_0^{1/2} \exp\left(\frac{(1-v_0)}{2v_0}\frac{\alpha_{j}^{2(t)}}{\tau^{2(t)}_j}\right)$
    \STATE update $w^{(t)}$ from its FCD: 
    $w^{(t)}|\cdot \sim \operatorname{Beta}\left(a_w + \sum_j^p I_1(\gamma^{(t)}_j),  b_w + \sum_j^p I_{v0}(\gamma^{(t)}_j)\right)$
    \IF{$\bm y$ is Gaussian} \STATE update ${\phi}^{(t)}$ from its FCD: 
    $\phi^{(t)}|\cdot \sim  \Gamma^{-1}\left(a_\phi + n/2,  b_\phi + \frac{\sum^n_i(y_i-\eta^{(t)}_i)^2}{2}\right) $ \ENDIF
\ENDFOR
\end{algorithmic}
\caption{MCMC sampler for peNMIG} \label{MCMCalgFCD}
\end{algorithm}

\pkg{spikeSlabGAM} uses the blockwise Gibbs sampler summarized in Algorithm~\ref{MCMCalgFCD} for MCMC inference. 
The sampler cyclically updates the nodes in Figure~\ref{parExpDAG}.  
The FCD for $\bm\alpha$ is based on the ``collapsed'' design matrix \mbox{$\bm X_\alpha = \bm
X \operatorname{blockdiag}(\bm\xi_1,\dots,\bm\xi_p)$}, while $\bm\xi$
is sampled based on a ``rescaled'' design matrix \mbox{$\bm X_\xi = \bm X
\operatorname{blockdiag}({\bm 1_d}_1,\dots,{\bm 1_d}_p)\bm\alpha$},
where $\bm 1_d$ is a $d \times 1$ vector of ones and $\bm X = \left[\bm X_u\; \bm B_1 \ldots  \bm B_p \right]$
is the concatenation of the designs for the different model terms (see \eqref{F:eta}). 
The full conditionals for $\bm\alpha$ and $\bm\xi$ for Gaussian responses are
given by
\begin{align}
\begin{split}
 \bm\alpha|\cdot &\sim N(\bm \mu_{\alpha}, \Sigma_{\alpha}) \text{ with}\\
 \bm \Sigma_{\alpha} &= \left(\frac{1}{\phi}\bm X_\alpha^\top\bm X_\alpha +
\operatorname{diag}\left(\bm \gamma \bm \tau^2\right)^{-1}\right)^{-1}, \; \bm\mu_j
=\frac{1}{\phi}\bm\Sigma_\alpha\bm X_\alpha^\top\bm y,  
\text{ and}\\
 \bm\xi|\cdot &\sim N(\bm \mu_{\xi}, \bm \Sigma_{\xi}) \text{ with}\\
 \bm \Sigma_{\xi} &= \left(\frac{1}{\phi}\bm X_\xi^\top\bm X_\xi +
\bm I \right)^{-1}; \; \bm\mu_j
=\bm\Sigma_\xi\left(\frac{1}{\phi}\bm X_\xi^\top\bm y + \bm m \right).  
\end{split}\label{F:FCDCoef} 
\end{align}

For non-Gaussian responses, we use penalized iteratively re-weighted least squares (P-IWLS)
proposals \citep{Lang:Brezger:2004} in a Metropolis-Hastings step
to sample $\bm \alpha$ and $\bm \xi$, i.e., Gaussian proposals are drawn from the quadratic Taylor
approximation of the logarithm of the intractable FCD. Because of the prohibitive computational cost for large $q$ and $p$ 
(and low acceptance rates for non-Gaussian response for high-dimensional IWLS proposals), 
neither $\bm \alpha$ nor $\bm \xi$ are updated all at once. Rather, both $\bm \alpha$ and $\bm \xi$ are split 
into $b_\alpha$ ($b_\xi$) update blocks that are updated sequentially conditional on the states of all 
other parameters. 

By default, starting values $\bm \beta^{(0)}$ are drawn randomly in three steps:  
First, 5 Fisher scoring steps with fixed, large hypervariances are performed to reach a viable region of the parameter space. 
Second, for each chain run in parallel, Gaussian noise is added to this preliminary $\bm \beta^{(0)}$, 
and third its constituting 
$p$ subvectors are scaled with
variance parameters $\gamma_j\tau_j^2\;( j=1,\dots,p)$ drawn from their priors.
this means that, for each of the parallel chains, some of the $p$ model terms are set
close to zero initially, and the remainder is in the vicinity of their respective ridge-penalized MLEs.  
Starting values for
$\bm\alpha^{(0)}$ and $\bm\xi^{(0)}$ are then computed via $
\alpha^{(0)}_j = d_j^{-1} \sum_i^{d_j} |\beta^{(0)}_{ji}|$ and $\bm \xi^{(0)}_j = \bm\beta^{(0)}_j / \alpha^{(0)}_j$.
Section 3.2 in \citet{Scheipl:2011} contains more details on the
sampler.         

\section{Using spikeSlabGAM}\label{using}

\subsection{Model specification and post-processing}\label{specification}
\pkg{spikeSlabGAM} uses the standard \proglang{R} formula syntax to specify models, with a slight twist:  
Every term in the model has to belong to one of the term types given 
in Table \ref{T:termtypes}. If a model formula contains ``raw'' terms not wrapped in one of 
these term type functions, the package will try to guess appropriate term types: For example, the formula \code{y }$\ftilde$\code{ x + f} with a numeric \code{x} and a factor \code{f} is expanded
into \code{y }$\ftilde$\code{ lin(x) + sm(x) + fct(f)} since the default is to model any numeric covariate 
as a smooth effect with a \code{lin()}-term parameterizing functions from the nullspace of its penalty
and an \code{sm()}-term parameterizing the penalized part. 
The model formula defines the candidate set of model terms that comprise the
model of maximal complexity under consideration. As of now, indicators $\bm\gamma$ are sampled without hierarchical constraints, 
i.e., an interaction effect can be included in the model even if the associated main effects or lower order interactions are not.

We generate some artificial data for a didactic example.
We draw $n=200$ observations from the following data generating process:
\begin{itemize}
\item covariates \code{sm1, sm2, noise2, noise3} are $\stackrel{\iid}{\sim} U[0,1]$,
\item covariates \code{f, noise4} are factors with 3 and 4 levels,
\item covariates \code{lin1, lin2, lin3} are $\stackrel{\iid}{\sim} N(0,1)$,
\item covariate \code{noise1} is collinear with \code{sm1}: $\mathtt{noise1} = \mathtt{sm1} + \bm e;\; e_i \stackrel{\iid}{\sim} N(0,1)$,
\item $\bm{\eta}  = f(\mathtt{sm1}) + f(\mathtt{sm2, f}) + 0.1 \cdot \mathtt{lin1} + 0.2\cdot\mathtt{lin2} + 0.3\cdot\mathtt{lin3}$
        (see Figures~\ref{fig:m1Sm1} and \ref{fig:m1Sm2f} for the shapes of the nonlinear effects $f(\mathtt{sm1})$ and $f(\mathtt{sm2, f})$),
\item  the response vector $\bm y = \bm{\eta} +  \tfrac{\operatorname{sd}(\bm \eta)}{\mathtt{snr}} \bm\epsilon$ is generated under signal-to-noise ratio $\mathtt{snr}=3$
with i.i.d.~$t_5$-distributed errors $\epsilon_i\;(i=1,\ldots,n)$.         
\end{itemize}
\begin{Schunk}
\begin{Sinput}
R> set.seed(1312424)
R> n <- 200
R> snr <- 3
R> sm1 <- runif(n)
R> fsm1 <- dbeta(sm1, 7, 3)/2
R> sm2 <- runif(n, 0, 1)
R> f <- gl(3, n/3)
R> ff <- as.numeric(f)/2
R> fsm2f <- ff + ff * sm2 + ((f == 1) * -dbeta(sm2, 6, 4) + 
+      (f == 2) * dbeta(sm2, 6, 9) + (f == 3) * dbeta(sm2, 
+      9, 6))/2
R> lin <- matrix(rnorm(n * 3), n, 3)
R> colnames(lin) <- paste("lin", 1:3, sep = "")
R> noise1 <- sm1 + rnorm(n)
R> noise2 <- runif(n)
R> noise3 <- runif(n)
R> noise4 <- sample(gl(4, n/4))
R> eta <- fsm1 + fsm2f + lin %*% c(0.1, 0.2, 0.3)
R> y <- eta + sd(eta)/snr * rt(n, df = 5)
R> d <- data.frame(y, sm1, sm2, f, lin, noise1, noise2, 
+      noise3, noise4)
\end{Sinput}
\end{Schunk}

We fit an additive model with all covariates as main effects and first-order interactions between the first 4 as potential model terms:
\begin{Schunk}
\begin{Sinput}
R> f1 <- y ~ (sm1 + sm2 + f + lin1)^2 + lin2 + lin3 + noise1 + 
+      noise2 + noise3 + noise4
\end{Sinput}
\end{Schunk}
The function \code{spikeSlabGAM} sets up the design matrices, calls the sampler and returns the results: 
\begin{Schunk}
\begin{Sinput}
R> m <- spikeSlabGAM(formula = f1, data = d)
\end{Sinput}
\end{Schunk}

The following output shows the first part of the \code{summary} of the fitted model.
Note that the numeric covariates have been split into \code{lin()}- and \code{sm()}-terms and that the factors
have been correctly identified as \code{fct()}-terms. The joint effect of the two numerical covariates \code{sm1} and \code{sm2}
has been decomposed into 8 components: the 4 marginal linear and smooth terms, their linear-linear interaction, two ``varying coefficient'' terms 
(i.e., linear-smooth interactions) and a smooth interaction surface. This decomposition can be helpful in constructing parsimonious models.
If a decomposition into marginal and joint effects is irrelevant or inappropriate, 
bivariate smooth terms can alternatively be specified with a \code{srf()}-term. 
\code{Mean posterior deviance} is $\tfrac{1}{T}\sum^T_t -2l(\bm y| \bm \eta^{(t)}, \phi^{(t)})$, the average of twice the negative 
log-likelihood of the observations over the saved MCMC iterations, the \code{null deviance} is twice the negative 
log-likelihood of an intercept model without covariates.
\begin{Schunk}
\begin{Sinput}
R> summary(m)
\end{Sinput}
\end{Schunk}
\begin{small}
\begin{Schunk}
\begin{Soutput}
Spike-and-Slab STAR for Gaussian data 

Model:
y ~ ((lin(sm1) + sm(sm1)) + (lin(sm2) + sm(sm2)) + fct(f) + (lin(lin1) + 
    sm(lin1)))^2 + (lin(lin2) + sm(lin2)) + (lin(lin3) + sm(lin3)) + 
    (lin(noise1) + sm(noise1)) + (lin(noise2) + sm(noise2)) + 
    (lin(noise3) + sm(noise3)) + fct(noise4) - lin(sm1):sm(sm1) - 
    lin(sm2):sm(sm2) - lin(lin1):sm(lin1)
200 observations; 257 coefficients in 37 model terms.

Prior:
    a[tau]     b[tau]       v[0]       a[w]       b[w] a[sigma^2] 
   5.0e+00    2.5e+01    2.5e-04    1.0e+00    1.0e+00    1.0e-04 
b[sigma^2] 
   1.0e-04 

MCMC:
Saved 1500 samples from 3 chain(s), each ran 2500 iterations after a
  burn-in of 100 ; Thinning: 5

                            
Null deviance:           704
Mean posterior deviance: 284
\end{Soutput}
\end{Schunk}
\begin{Schunk}
\begin{Soutput}


Marginal posterior inclusion probabilities and term importance:
                   P(gamma=1)     pi dim    
u                          NA     NA   1    
lin(sm1)                1.000  0.097   1 ***
sm(sm1)                 1.000  0.065   8 ***
lin(sm2)                0.998  0.028   1 ***
sm(sm2)                 0.971  0.015   8 ***
fct(f)                  1.000  0.576   2 ***
lin(lin1)               0.082 -0.002   1    
sm(lin1)                0.030  0.001   9    
lin(lin2)               0.997  0.029   1 ***
sm(lin2)                0.066  0.001   9    
lin(lin3)               1.000  0.043   1 ***
sm(lin3)                0.037  0.000   9    
lin(noise1)             0.059  0.002   1    
sm(noise1)              0.036  0.000   9    
lin(noise2)             0.017  0.000   1    
sm(noise2)              0.032  0.000   8    
lin(noise3)             0.025  0.000   1    
sm(noise3)              0.034  0.000   8    
fct(noise4)             0.081  0.001   3    
lin(sm1):lin(sm2)       0.022  0.000   1    
lin(sm1):sm(sm2)        0.064  0.000   7    
lin(sm1):fct(f)         0.115 -0.003   2    
lin(sm1):lin(lin1)      0.020  0.000   1    
lin(sm1):sm(lin1)       0.056  0.000   7    
sm(sm1):lin(sm2)        0.044  0.000   7    
sm(sm1):sm(sm2)         0.110 -0.001  27    
sm(sm1):fct(f)          0.061  0.000  13    
sm(sm1):lin(lin1)       0.050  0.000   7    
sm(sm1):sm(lin1)        0.061  0.000  28    
lin(sm2):fct(f)         1.000  0.054   2 ***
lin(sm2):lin(lin1)      0.018  0.000   1    
lin(sm2):sm(lin1)       0.055  0.000   8    
sm(sm2):fct(f)          1.000  0.092  13 ***
sm(sm2):lin(lin1)       0.062  0.000   7    
sm(sm2):sm(lin1)        0.209  0.000  28    
fct(f):lin(lin1)        0.102  0.000   2    
fct(f):sm(lin1)         0.179  0.001  14    
*:P(gamma=1)>.25 **:P(gamma=1)>.5 ***:P(gamma=1)>.9
\end{Soutput}
\end{Schunk}
\end{small}
\begin{figure}[!htbp]
\begin{small}
\begin{Schunk}
\begin{Soutput}
Posterior model probabilities (inclusion threshold = 0.5 ):
\end{Soutput}
\begin{Soutput}
                       1     2     3     4     5     6     7     8
prob.:             0.307 0.059 0.045 0.028 0.022  0.02 0.017 0.013
lin(sm1)               x     x     x     x     x     x     x     x
sm(sm1)                x     x     x     x     x     x     x     x
lin(sm2)               x     x     x     x     x     x     x     x
sm(sm2)                x     x     x     x     x     x     x     x
fct(f)                 x     x     x     x     x     x     x     x
lin(lin1)                                            x            
sm(lin1)                                                          
lin(lin2)              x     x     x     x     x     x     x     x
sm(lin2)                                                          
lin(lin3)              x     x     x     x     x     x     x     x
sm(lin3)                                                          
lin(noise1)                                                       
sm(noise1)                                                        
lin(noise2)                                                       
sm(noise2)                                                        
lin(noise3)                                                       
sm(noise3)                                                        
fct(noise4)                                                       
lin(sm1):lin(sm2)                                                 
lin(sm1):sm(sm2)                                                 x
lin(sm1):fct(f)                                x                  
lin(sm1):lin(lin1)                                                
lin(sm1):sm(lin1)                                                 
sm(sm1):lin(sm2)                                                  
sm(sm1):sm(sm2)                                            x      
sm(sm1):fct(f)                                                    
sm(sm1):lin(lin1)                                                 
sm(sm1):sm(lin1)                                                  
lin(sm2):fct(f)        x     x     x     x     x     x     x     x
lin(sm2):lin(lin1)                                                
lin(sm2):sm(lin1)                                                 
sm(sm2):fct(f)         x     x     x     x     x     x     x     x
sm(sm2):lin(lin1)                                                 
sm(sm2):sm(lin1)             x                                    
fct(f):lin(lin1)                         x                        
fct(f):sm(lin1)                    x                              
cumulative:        0.307 0.367 0.412  0.44 0.462 0.482 0.499 0.513
\end{Soutput}
\end{Schunk}
\end{small}
\caption{Excerpt of the second part of the output returned by \texttt{summary.spikeSlabGAM}, which tabulates the  
configurations of $P(\gamma_j=1)>.5$ with highest posterior probability. In the example,
the posterior is very concentrated in the true model without \code{lin1}, which has a posterior 
probability of 0.31. The correct model 
that additionally includes \code{lin1} (column 6)
has a posterior probability of 0.02.}
\label{sum12}
\end{figure}
In most applications, the primary focus will be on the marginal posterior inclusion probabilities \code{P(gamma = 1)}, 
given along with a measure of term importance \code{pi} and the size of the associated coefficient batch \code{dim}. 
\code{pi} is defined as $\pi_j = \bm{\bar{\eta}}_j^\top\bm{\bar{\eta}}_{-1}/\bm{\bar{\eta}}_{-1}^T\bm{\bar{\eta}}_{-1}$, 
where $\bm{\bar{\eta}}_j$ is the posterior expectation of the linear predictor associated with the $j^{th}$ term, and $\bm{\bar{\eta}}_{-1}$ 
is the linear predictor minus the intercept. Since $\sum^p_j\pi_j=1$,  the \code{pi} values provide a rough percentage decomposition 
of the sum of squares of the (non-constant) linear predictor \citep{Gu:1992}. Note that they can assume negative values as well for terms
whose contributions to the linear predictor $\bm{\bar{\eta}}_j$ are negatively correlated with the remainder of the (non-constant)
linear predictor $\bm{\bar{\eta}}_{-1} - \bm{\bar{\eta}}_j$. 
The summary shows that almost all true effects have a high posterior inclusion probability (i.e.,~\code{lin()} for \code{lin2, lin3}; 
\code{lin(),sm()} for \code{sm1, sm2}; \code{fct(f)}; and the interaction terms between \code{sm2} and \code{f}). 
All the terms associated with noise variables and the superfluous smooth terms for \code{lin1, lin2, lin3} as well as the
superfluous interaction terms have a very low posterior inclusion probability.  The small linear influence of \code{lin1}
has not been recovered. 

Figure~\ref{sum12} shows an excerpt from the second part of the \code{summary} output, which summarizes
the posterior of the vector of inclusion indicators $\bm \gamma$. 
The table shows the different configurations
of $P(\gamma_j =1)>.5, j=1,\ldots,p$ sorted by relative frequency, i.e., the models visited by the sampler sorted by decreasing posterior support. For this simulated data,
the posterior is concentrated strongly on the (almost) true model missing the small linear effect of \code{lin1}.

\subsection{Visualization}\label{visualization}
\pkg{spikeSlabGAM} offers automated visualizations for model terms and their interactions, implemented with 
\pkg{ggplot2} \citep{ggplot2}. 
By default, the posterior mean of the linear predictor associated with each covariate 
(or combination of covariates if the model contains interactions) along with (pointwise) 80\% 
credible intervals is shown. 
\setkeys{Gin}{width=6in}
\begin{figure}[!htbp]\begin{center}
\tikzset{font=\scriptsize}
\begin{Schunk}
\begin{Sinput}
R> plot(m)
\end{Sinput}
\end{Schunk}
\beginpgfgraphicnamed{UsingSpikeSlabGAM-plotm1}
\input{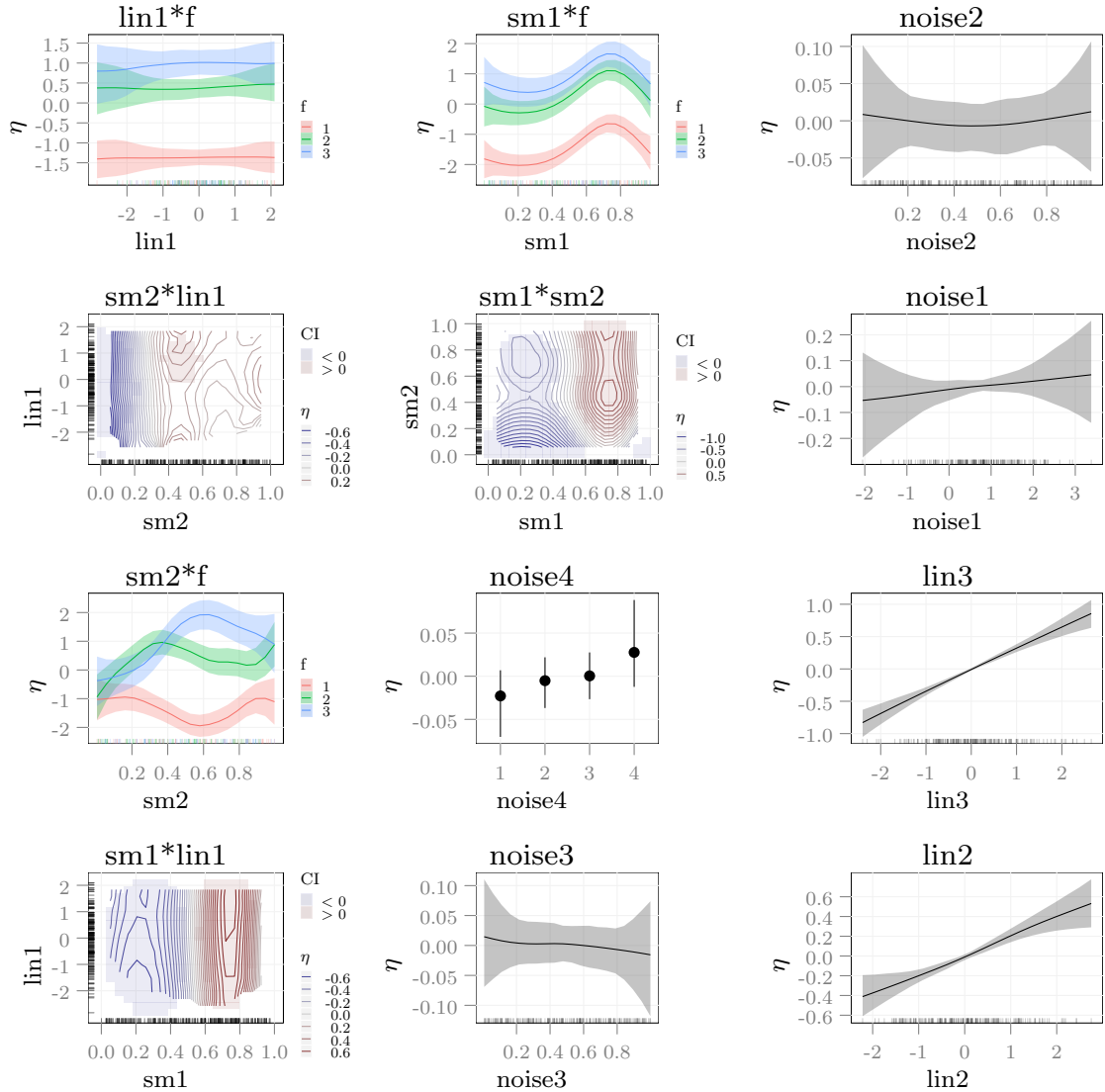}
\endpgfgraphicnamed
\caption{Posterior means and pointwise 80\% credible intervals for \texttt{m1}. 
Interaction surfaces of two numerical covariates 
are displayed as color coded contour plots, with regions in which the credible interval does not overlap zero marked 
in blue ($\eta < 0$) or red ($\eta > 0$). Each panel contains a marginal rug plot that shows where the observations are located. 
Note that the default behavior of
\texttt{plot.spikeSlabGAM} is to cumulate all terms associated with a covariate or covariate combination. 
In this example, the joint effects of the first 4 covariates \code{sm1, sm2, f} and \code{lin1} and the sums of the
\texttt{lin}- and \texttt{sm}-terms associated with \code{lin2, lin3, noise1, noise2} and \code{noise3}
are displayed. All effects of the noise variables are $\approx 0$, note the different scales on the vertical axes.
Vertical axes can be forced to the same range by setting option \code{commonEtaScale}.}
\label{fig:m1}
\end{center}\end{figure}
Figure~\ref{fig:m1} shows the estimated effects for \code{m1}.  

\begin{figure}[!htbp]\begin{center}
\tikzset{font=\small}
\begin{Schunk}
\begin{Sinput}
R> plot(m, labels = c("lin(sm1)", "sm(sm1)"), cumulative = FALSE)
R> trueFsm1 <- data.frame(truth = fsm1 - mean(fsm1), sm1 = sm1)
R> plot(m, labels = "sm(sm1)", ggElems = list(geom_line(aes(x = sm1, 
+      y = truth), data = trueFsm1, linetype = 2)))
\end{Sinput}
\end{Schunk}
\beginpgfgraphicnamed{UsingSpikeSlabGAM-plotm1sm1}
\input{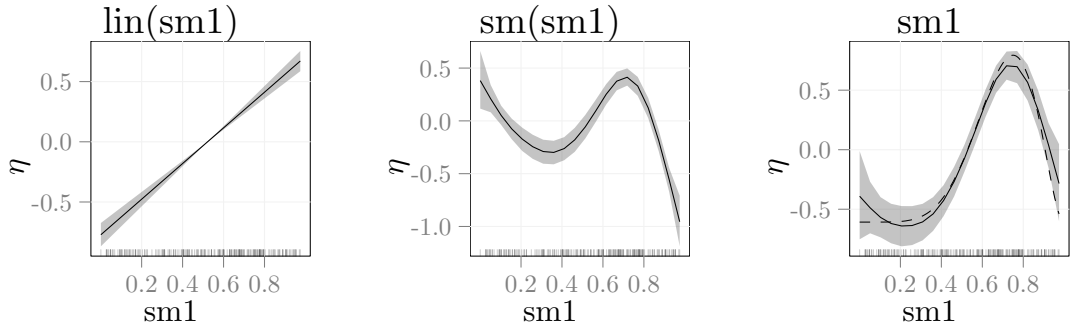}
\endpgfgraphicnamed
\caption{Posterior means and pointwise 80\% credible intervals for $f(\mathtt{sm1})$ in 
\texttt{m1}. Left and middle panel show the separate \texttt{lin()}- and \texttt{sm()}-terms returned by the first call to \code{plot}, 
right panel shows their sum. True shape of $f(\mathtt{sm1})$ added as a dashed line with the \texttt{ggElems} option of \texttt{plot.spikeSlabGAM}.}
\label{fig:m1Sm1}
\end{center}\end{figure}

\begin{figure}[!htbp]\begin{center}
\tikzset{font=\small}
\begin{Schunk}
\begin{Sinput}
R> trueFsm2f <- data.frame(truth = fsm2f - mean(fsm2f), 
+      sm2 = sm2, f = f)
R> plot(m, labels = "sm(sm2):fct(f)", ggElems = list(geom_line(aes(x = sm2, 
+      y = truth, colour = f), data = trueFsm2f, linetype = 2)))
\end{Sinput}
\end{Schunk}
\beginpgfgraphicnamed{UsingSpikeSlabGAM-plotm1sm2f}
\input{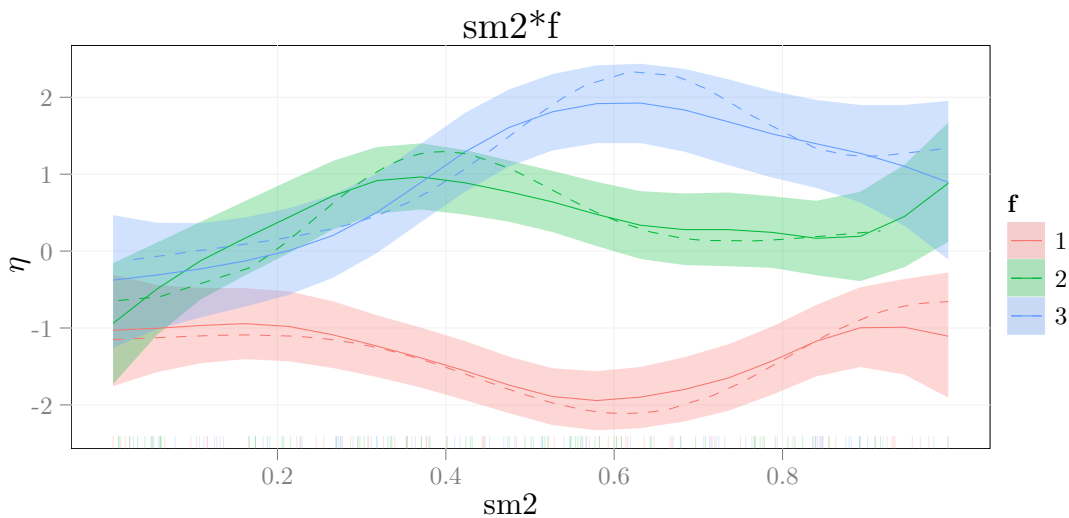}
\endpgfgraphicnamed
\caption{Posterior means and pointwise 80\% credible intervals for $f(\mathtt{sm2}, \mathtt{f})$ in 
\texttt{m1}. True shape of $f(\mathtt{sm2}|\mathtt{f})$ added as dashed line for each level of \texttt{f}.}
\label{fig:m1Sm2f}
\end{center}\end{figure}

Plots for specific terms can be requested with the \code{label} argument, Figures~\ref{fig:m1Sm1} and \ref{fig:m1Sm2f}
show code snippets and their output for $f(\mathtt{sm1})$ and $f(\mathtt{sm2, f})$. 
The fits are quite close to the truth despite the heavy-tailed errors and the many noise terms included in the model. 
Full disclosure: The code used to render Figures~\ref{fig:m1Sm1} and \ref{fig:m1Sm2f} is a little more intricate
than the code snippets we show, but the additional code only affects details (font and margin sizes and the arrangement of the 
panels).  
\setkeys{Gin}{width=6in}

\setkeys{Gin}{width=6in}

\subsection{Assessing convergence}\label{convergence}
\sloppy
\pkg{spikeSlabGAM} uses the convergence diagnostics implemented in \pkg{R2WinBUGS} \citep{R2WinBUGS}. 
The function \code{ssGAM2Bugs()} converts the posterior samples for a \code{spikeSlabGAM}-object into a \code{bugs}-object, for which graphical and numerical convergence diagnostics
are available via \code{plot} and \code{print}. Note that not all cases of non-convergence should be considered problematic, e.g., if one of the 
chains samples from a different part of the model space than the others, but has converged on that part of the parameter space.   
\fussy

\subsection{Pima indian diabetes}\label{pima}
We use the time-honored Pima Indian Diabetes dataset as an example for real non-gaussian data:
This dataset from the UCI repository \citep{uci} is provided in package \pkg{mlbench} \citep{mlbench} as \code{PimaIndiansDiabetes2}.
We remove two columns with a large number of missing values and
use the complete measurements of the remaining 7 covariates and the response (diabetes Yes/No) for 524 women
to estimate the model. 
We set aside 200 observations as a test set:
\begin{Schunk}
\begin{Sinput}
R> data("PimaIndiansDiabetes2", package = "mlbench")
R> pimaDiab <- na.omit(PimaIndiansDiabetes2[, -c(4, 5)])
R> pimaDiab <- within(pimaDiab, {
+      diabetes <- 1 * (diabetes == "pos")
+  })
R> set.seed(1109712439)
R> testInd <- sample(1:nrow(pimaDiab), 200)
R> pimaDiabTrain <- pimaDiab[-testInd, ]
\end{Sinput}
\end{Schunk}
Note that \code{spikeSlabGAM()} always expects a dataset without any missing values and responses between 0 and 1 for binomial models.

We increase the length of the burn-in phase for each chain from 100 to 500 iterations and 
run 8 parallel chains for an additive main effects model 
(if either \pkg{multicore} \citep{multicore} or \pkg{snow} \citep{snow} are installed, the chains will be run in parallel): 
\begin{Schunk}
\begin{Sinput}
R> mcmc <- list(nChains = 8, chainLength = 1000, burnin = 500, 
+      thin = 5)
R> m0 <- spikeSlabGAM(diabetes ~ pregnant + glucose + pressure + 
+      mass + pedigree + age, family = "binomial", data = pimaDiabTrain, 
+      mcmc = mcmc)
\end{Sinput}
\end{Schunk}
We compute the posterior predictive means for the test set, and request a summary of the fitted model: 
\begin{Schunk}
\begin{Sinput}
R> pr0 <- predict(m0, newdata = pimaDiab[testInd, ])
R> print(summary(m0), printModels = FALSE)
\end{Sinput}
\begin{Soutput}
Spike-and-Slab STAR for Binomial data 

Model:
diabetes ~ (lin(pregnant) + sm(pregnant)) + (lin(glucose) + sm(glucose)) + 
    (lin(pressure) + sm(pressure)) + (lin(mass) + sm(mass)) + 
    (lin(pedigree) + sm(pedigree)) + (lin(age) + sm(age))
524 observations; 58 coefficients in 13 model terms.

Prior:
 a[tau]  b[tau]    v[0]    a[w]    b[w] 
5.0e+00 2.5e+01 2.5e-04 1.0e+00 1.0e+00 

MCMC:
Saved 8000 samples from 8 chain(s), each ran 5000 iterations after a
  burn-in of 500 ; Thinning: 5
P-IWLS acceptance rates: 0.92 for alpha; 0.64 for xi.
                            
Null deviance:           676
Mean posterior deviance: 474

Marginal posterior inclusion probabilities and term importance:
              P(gamma=1)     pi dim    
u                     NA     NA   1    
lin(pregnant)      0.120  0.009   1    
sm(pregnant)       0.018  0.000   8    
lin(glucose)       1.000  0.517   1 ***
sm(glucose)        0.020  0.000   9    
lin(pressure)      0.020 -0.001   1    
sm(pressure)       0.011  0.000   9    
lin(mass)          1.000  0.237   1 ***
sm(mass)           0.628  0.035   9  **
lin(pedigree)      0.024  0.001   1    
sm(pedigree)       0.370 -0.002   8   *
lin(age)           0.511  0.039   1  **
sm(age)            0.896  0.164   8  **
*:P(gamma=1)>.25 **:P(gamma=1)>.5 ***:P(gamma=1)>.9
\end{Soutput}
\end{Schunk}
\code{spikeSlabGAM} selects nonlinear effects for \code{age} and \code{mass} and a linear trend in \code{glucose}.
The nonlinear effect of \code{pedigree} has fairly weak support.
\pkg{mboost}\code{::gamboost} ranks the variables very similarly, based on the relative selection frequencies of
the associated baselearners:
\begin{Schunk}
\begin{Sinput}
R> b <- gamboost(as.factor(diabetes) ~ pregnant + glucose + 
+      pressure + mass + pedigree + age, family = Binomial(), 
+      data = pimaDiabTrain)[300]
R> aic <- AIC(b, method = "classical")
R> prB <- predict(b[mstop(aic)], newdata = pimaDiab[testInd, 
+      ])
\end{Sinput}
\end{Schunk}
\begin{Schunk}
\begin{Sinput}
R> summary(b[mstop(aic)])$selprob
\end{Sinput}
\begin{Soutput}
    bbs(mass, df = dfbase)  bbs(glucose, df = dfbase) 
                  0.290323                   0.266129 
     bbs(age, df = dfbase) bbs(pedigree, df = dfbase) 
                  0.209677                   0.120968 
bbs(pregnant, df = dfbase) bbs(pressure, df = dfbase) 
                  0.072581                   0.040323 
\end{Soutput}
\end{Schunk}
Finally, we compare the deviance on the test set for the two fitted models:
\begin{Schunk}
\begin{Sinput}
R> dev <- function(y, p) {
+      -2 * sum(dbinom(x = y, size = 1, prob = p, log = T))
+  }
R> c(spikeSlabGAM = dev(pimaDiab[testInd, "diabetes"], pr0), 
+      gamboost = dev(pimaDiab[testInd, "diabetes"], plogis(prB)))
\end{Sinput}
\begin{Soutput}
spikeSlabGAM     gamboost 
      180.51       194.79 
\end{Soutput}
\end{Schunk}
So it seems like \code{spikeSlabGAM}'s model averaged predictions are a little more accurate 
than the predictions returned by \code{gamboost} in this case. 

We can check the sensitivity of the results to the hyperparameters and refit the 
model with a larger $v_0$ to see if/how the results change:
\begin{Schunk}
\begin{Sinput}
R> hyper1 <- list(gamma = c(v0 = 0.005))
R> m1 <- spikeSlabGAM(diabetes ~ pregnant + glucose + pressure + 
+      mass + pedigree + age, family = "binomial", data = pimaDiabTrain, 
+      mcmc = mcmc, hyperparameters = hyper1)
R> pr1 <- predict(m1, newdata = pimaDiab[testInd, ])
\end{Sinput}
\end{Schunk}
\begin{Schunk}
\begin{Sinput}
R> print(summary(m1), printModels = FALSE)
\end{Sinput}
\begin{Soutput}
Spike-and-Slab STAR for Binomial data 

Model:
diabetes ~ (lin(pregnant) + sm(pregnant)) + (lin(glucose) + sm(glucose)) + 
    (lin(pressure) + sm(pressure)) + (lin(mass) + sm(mass)) + 
    (lin(pedigree) + sm(pedigree)) + (lin(age) + sm(age))
524 observations; 58 coefficients in 13 model terms.

Prior:
a[tau] b[tau]   v[0]   a[w]   b[w] 
 5.000 25.000  0.005  1.000  1.000 

MCMC:
Saved 8000 samples from 8 chain(s), each ran 5000 iterations after a
  burn-in of 500 ; Thinning: 5
P-IWLS acceptance rates: 0.85 for alpha; 0.64 for xi.
                            
Null deviance:           676
Mean posterior deviance: 458

Marginal posterior inclusion probabilities and term importance:
              P(gamma=1)     pi dim    
u                     NA     NA   1    
lin(pregnant)      0.069  0.003   1    
sm(pregnant)       0.080 -0.001   8    
lin(glucose)       1.000  0.447   1 ***
sm(glucose)        0.079  0.000   9    
lin(pressure)      0.117 -0.010   1    
sm(pressure)       0.067  0.000   9    
lin(mass)          1.000  0.240   1 ***
sm(mass)           0.973  0.067   9 ***
lin(pedigree)      0.148  0.008   1    
sm(pedigree)       0.298  0.004   8   *
lin(age)           0.961  0.089   1 ***
sm(age)            0.996  0.153   8 ***
*:P(gamma=1)>.25 **:P(gamma=1)>.5 ***:P(gamma=1)>.9
\end{Soutput}
\begin{Sinput}
R> (dev(pimaDiab[testInd, "diabetes"], pr1))
\end{Sinput}
\begin{Soutput}
[1] 177.85
\end{Soutput}
\end{Schunk}
The selected terms are very similar, and the prediction is slightly more accurate 
(predictive deviance for \code{m0} was 180.51).
Note that, due to the multi-modality of the target posterior, stable estimation of precise posterior inclusion and
model probabilities requires more parallel chains than were used in this example. 

\section{Summary}
We have presented a novel approach for Bayesian variable selection, model choice, and regularized estimation in 
(geo-)additive mixed models for Gaussian, binomial, and Poisson responses implemented in \pkg{spikeSlabGAM}.
The package uses the established \proglang{R} formula syntax so that complex models can be specified very concisely.
It features powerful and user friendly visualizations of the fitted models. 
Major features of the software have been demonstrated on an example with artificial data with t-distributed errors
and on the Pima Indians Diabetes data set. 
In future work, the author plans to add capabilities for "always included" semiparametric terms and for 
sampling the inclusion indicators under hierarchical constraints, 
i.e., never including an interaction if the associated main effects are excluded. 

\section*{Acknowledgements} 
Discussions with and feedback from Thomas Kneib, Ludwig
Fahrmeir and Simon N. Wood were enlightening and inspiring. 
Susanne Konrath shared an office with the author and his slovenly ways without complaint.
Brigitte Maxa fought like a lioness to make sure the author kept getting a paycheck.
Financial support from the German Science Foundation (grant FA 128/5-1) is gratefully acknowledged.

\bibliography{UsingSpikeSlabGAM}

\begin{thebibliography}{36}
\newcommand{\enquote}[1]{``#1''}
\providecommand{\natexlab}[1]{#1}
\providecommand{\url}[1]{\texttt{#1}}
\providecommand{\urlprefix}{URL }
\expandafter\ifx\csname urlstyle\endcsname\relax
  \providecommand{\doi}[1]{doi:\discretionary{}{}{}#1}\else
  \providecommand{\doi}{doi:\discretionary{}{}{}\begingroup
  \urlstyle{rm}\Url}\fi
\providecommand{\eprint}[2][]{\url{#2}}

\bibitem[{Bae and Mallick(2004)}]{Bae:Mallick:2004}
Bae K, Mallick B (2004).
\newblock \enquote{Gene Selection Using a Two-Level Hierarchical Bayesian
  Model.}
\newblock \emph{Bioinformatics}, \textbf{20}(18), 3423--3430.

\bibitem[{Baglama and Reichel(2006)}]{Baglama:Reichel:2006}
Baglama J, Reichel L (2006).
\newblock \enquote{Augmented Implicitly Restarted Lanczos Bidiagonalization
  Methods.}
\newblock \emph{SIAM Journal on Scientific Computing}, \textbf{27}(1), 19--42.

\bibitem[{Brezger \emph{et~al.}(2005)Brezger, Kneib, and Lang}]{bayesx}
Brezger A, Kneib T, Lang S (2005).
\newblock \enquote{\proglang{BayesX}: Analyzing Bayesian Structural Additive
  Regression Models.}
\newblock \emph{Journal of Statistical Software}, \textbf{14}(11).

\bibitem[{Carvalho \emph{et~al.}(2010)Carvalho, Polson, and
  Scott}]{Carvalho:2010}
Carvalho C, Polson N, Scott J (2010).
\newblock \enquote{The Horseshoe Estimator for Sparse Signals.}
\newblock \emph{Biometrika}, \textbf{97}(2), 465--480.

\bibitem[{Cottet \emph{et~al.}(2008)Cottet, Kohn, and
  Nott}]{Cottet:Kohn:Nott:2008}
Cottet R, Kohn R, Nott D (2008).
\newblock \enquote{Variable Selection and Model Averaging in Semiparametric
  Overdispersed Generalized Linear Models.}
\newblock \emph{Journal of the American Statistical Association},
  \textbf{103}(482), 661--671.

\bibitem[{Fahrmeir \emph{et~al.}(2010)Fahrmeir, Kneib, and
  Konrath}]{Fahr:Kneib:Konrath:2010}
Fahrmeir L, Kneib T, Konrath S (2010).
\newblock \enquote{Bayesian Regularisation in Structured Additive Regression: a
  Unifying Perspective on Shrinkage, Smoothing and Predictor Selection.}
\newblock \emph{Statistics and Computing}, \textbf{20}(2), 203--219.

\bibitem[{Fahrmeir \emph{et~al.}(2004)Fahrmeir, Kneib, and
  Lang}]{Fahr:Kneib:Lang:2004}
Fahrmeir L, Kneib T, Lang S (2004).
\newblock \enquote{Penalized Structured Additive Regression for Space-Time
  Data: a Bayesian Perspective.}
\newblock \emph{Statistica Sinica}, \textbf{14}, 731--761.

\bibitem[{Fr\"uhwirth-Schnatter and Wagner(2010)}]{Fruehwirt:Wagner:2010}
Fr\"uhwirth-Schnatter S, Wagner H (2010).
\newblock \enquote{Bayesian Variable Selection for Random Intercept Modelling
  of Gaussian and Non-Gaussian Data.}
\newblock In J~Bernardo, M~Bayarri, JO~Berger, AP~Dawid, D~Heckerman, AFM
  Smith, M~West (eds.), \emph{Bayesian Statistics 9}. Oxford University Press.

\bibitem[{Gelman \emph{et~al.}(2008)Gelman, Van~Dyk, Huang, and
  Boscardin}]{Gelman:2008}
Gelman A, Van~Dyk D, Huang Z, Boscardin J (2008).
\newblock \enquote{Using Redundant Parameterizations to Fit Hierarchical
  Models.}
\newblock \emph{Journal of Computational and Graphical Statistics},
  \textbf{17}(1), 95--122.

\bibitem[{George and McCulloch(1993)}]{George:McCulloch:1993}
George E, McCulloch R (1993).
\newblock \enquote{Variable Selection via Gibbs Sampling.}
\newblock \emph{Journal of the American Statistical Association},
  \textbf{88}(423), 881--889.

\bibitem[{Gu(1992)}]{Gu:1992}
Gu C (1992).
\newblock \enquote{Diagnostics for Nonparametric Regression Models with
  Additive Terms.}
\newblock \emph{Journal of the American Statistical Association},
  \textbf{87}(420), 1051--1058.

\bibitem[{Han and Carlin(2001)}]{Han:Carlin:2001}
Han C, Carlin B (2001).
\newblock \enquote{Markov Chain Monte Carlo Methods for Computing Bayes
  Factors.}
\newblock \emph{Journal of the American Statistical Association},
  \textbf{96}(455), 1122--1132.

\bibitem[{Hothorn \emph{et~al.}(2010)Hothorn, Buehlmann, Kneib, Schmid, and
  Hofner}]{mboost}
Hothorn T, Buehlmann P, Kneib T, Schmid M, Hofner B (2010).
\newblock \emph{\pkg{mboost}: Model-Based Boosting}.
\newblock \proglang{R} package version 2.0-7.

\bibitem[{Ishwaran and Rao(2005)}]{Ishwaran:2005}
Ishwaran H, Rao J (2005).
\newblock \enquote{Spike and Slab Variable Selection: Frequentist and Bayesian
  Strategies.}
\newblock \emph{The Annals of Statistics}, \textbf{33}(2), 730--773.

\bibitem[{Kneib(2006)}]{Kneib:06}
Kneib T (2006).
\newblock \emph{Mixed Model Based Inference in Structured Additive Regression}.
\newblock Dr. Hut Verlag.
\newblock \urlprefix\url{http://edoc.ub.uni-muenchen.de/archive/00005011/}.

\bibitem[{Lang and Brezger(2004)}]{Lang:Brezger:2004}
Lang S, Brezger A (2004).
\newblock \enquote{Bayesian P-Splines.}
\newblock \emph{Journal of Computational and Graphical Statistics},
  \textbf{13}(1), 183--212.

\bibitem[{Leisch and Dimitriadou(2010)}]{mlbench}
Leisch F, Dimitriadou E (2010).
\newblock \emph{\pkg{mlbench}: Machine Learning Benchmark Problems}.
\newblock \proglang{R} package version 2.1-0.

\bibitem[{Lewis(2009)}]{irlba}
Lewis B (2009).
\newblock \emph{\pkg{irlba}: Fast Partial SVD by Implicitly-Restarted Lanczos
  Bidiagonalization}.
\newblock \proglang{R} package version 0.1.1,
  \urlprefix\url{http://www.rforge.net/irlba/}.

\bibitem[{Lin and Zhang(2006)}]{Lin:Zhang:2006}
Lin Y, Zhang H (2006).
\newblock \enquote{{Component selection and smoothing in multivariate
  nonparametric regression}.}
\newblock \emph{The Annals of Statistics}, \textbf{34}(5), 2272--2297.

\bibitem[{Meng and van Dyk(1997)}]{Meng:Dyk:1997}
Meng X, van Dyk D (1997).
\newblock \enquote{The EM Algorithm--an Old Folk-Song Sung to a Fast New Tune.}
\newblock \emph{Journal of the Royal Statistical Society B}, \textbf{59}(3),
  511--567.

\bibitem[{Mitchell and Beauchamp(1988)}]{Mitchell:Beauchamp:1988}
Mitchell T, Beauchamp J (1988).
\newblock \enquote{Bayesian Variable Selection in Linear Regression.}
\newblock \emph{Journal of the American Statistical Association},
  \textbf{83}(404), 1023--1032.

\bibitem[{Newman \emph{et~al.}(1998)Newman, Hettich, Blake, and Merz}]{uci}
Newman D, Hettich S, Blake C, Merz C (1998).
\newblock \enquote{UCI Repository of machine learning databases.}
\newblock \urlprefix\url{http://www.ics.uci.edu/~mlearn/MLRepository.html}.

\bibitem[{Polson and Scott(2010)}]{Polson:Scott:2010}
Polson N, Scott J (2010).
\newblock \enquote{Shrink Globally, Act Locally: Sparse Bayesian Regularization
  and Prediction.}
\newblock In J~Bernardo, M~Bayarri, JO~Berger, AP~Dawid, D~Heckerman, AFM
  Smith, M~West (eds.), \emph{Bayesian Statistics 9}. Oxford University Press.

\bibitem[{{\proglang{R} Development Core Team}(2010)}]{R}
{\proglang{R} Development Core Team} (2010).
\newblock \emph{\proglang{R}: A Language and Environment for Statistical
  Computing}.
\newblock \proglang{R} Foundation for Statistical Computing, Vienna, Austria.
\newblock \urlprefix\url{http://www.R-project.org}.

\bibitem[{Reich \emph{et~al.}(2009)Reich, Storlie, and
  Bondell}]{Reich:Storlie:Bondell:2009}
Reich B, Storlie C, Bondell H (2009).
\newblock \enquote{Variable Selection in Bayesian Smoothing Spline ANOVA
  Models: Application to Deterministic Computer Codes.}
\newblock \emph{Technometrics}, \textbf{51}(2), 110.

\bibitem[{Rue and Held(2005)}]{Rue:Held:2005}
Rue H, Held L (2005).
\newblock \emph{Gaussian Markov Random Fields: Theory and Applications}.
\newblock Chapman \& Hall.

\bibitem[{Scheipl(2010)}]{Scheipl:2010}
Scheipl F (2010).
\newblock \enquote{Normal-Mixture-of-Inverse-Gamma Priors for Bayesian
  Regularization and Model Selection in Generalized Additive Models.}
\newblock \emph{Technical Report~84}, Department of Statistics, LMU München.
\newblock \urlprefix\url{http://epub.ub.uni-muenchen.de/11785/}.

\bibitem[{Scheipl(2011)}]{Scheipl:2011}
Scheipl F (2011).
\newblock \emph{Bayesian Regularization and Model Choice in Structured Additive
  Regression}.
\newblock Dr. Hut Verlag.
\newblock \urlprefix\url{http://edoc.ub.uni-muenchen.de/13005/}.

\bibitem[{Sturtz \emph{et~al.}(2005)Sturtz, Ligges, and Gelman}]{R2WinBUGS}
Sturtz S, Ligges U, Gelman A (2005).
\newblock \enquote{\pkg{R2WinBUGS}: A Package for Running WinBUGS from R.}
\newblock \emph{Journal of Statistical Software}, \textbf{12}(3), 1--16.

\bibitem[{Tierney \emph{et~al.}(2010)Tierney, Rossini, Li, and
  Sevcikova}]{snow}
Tierney L, Rossini A, Li N, Sevcikova H (2010).
\newblock \emph{\pkg{snow}: {S}imple {N}etwork {O}f {W}orkstations}.
\newblock \proglang{R} package version 0.3-3.

\bibitem[{Urbanek(2010)}]{multicore}
Urbanek S (2010).
\newblock \emph{\pkg{multicore}: Parallel Processing of R Code on Machines with
  Multiple Cores or CPUs}.
\newblock \proglang{R} package version 0.1-3,
  \urlprefix\url{http://www.rforge.net/multicore/}.

\bibitem[{Wahba \emph{et~al.}(1995)Wahba, Wang, Gu, Klein, and
  Klein}]{Wahba:Wang:Gu:1995}
Wahba G, Wang Y, Gu C, Klein R, Klein B (1995).
\newblock \enquote{Smoothing Spline ANOVA for Exponential Families, with
  Application to the Wisconsin Epidemiological Study of Diabetic Retinopathy.}
\newblock \emph{The Annals of Statistics}, \textbf{23}(6), 1865--1895.

\bibitem[{Wickham(2009)}]{ggplot2}
Wickham H (2009).
\newblock \emph{\pkg{ggplot2}: Elegant Graphics for Data Analysis}.
\newblock Springer-Verlag.
\newblock \urlprefix\url{http://had.co.nz/ggplot2/book}.

\bibitem[{Wood(2006)}]{Wood:2006}
Wood S (2006).
\newblock \emph{Generalized Additive Models: an Introduction with R}.
\newblock CRC Press.

\bibitem[{Wood \emph{et~al.}(2002)Wood, Kohn, Shively, and
  Jiang}]{Wood:Kohn:2002}
Wood S, Kohn R, Shively T, Jiang W (2002).
\newblock \enquote{Model Selection in Spline Nonparametric Regression.}
\newblock \emph{Journal of the Royal Statistical Society B}, \textbf{64}(1),
  119--139.

\bibitem[{Yau \emph{et~al.}(2003)Yau, Kohn, and Wood}]{Yau:Kohn:Wood:2003}
Yau P, Kohn R, Wood S (2003).
\newblock \enquote{Bayesian Variable Selection and Model Averaging in
  High-Dimensional Multinomial Nonparametric Regression.}
\newblock \emph{Journal of Computational and Graphical Statistics},
  \textbf{12}(1), 23--54.

\end{thebibliography}

\end{document}